\documentclass{article}
\usepackage{amsmath}
\usepackage{amssymb}
\usepackage{authblk}

\usepackage{lineno}

\usepackage{graphicx}
\usepackage{dcolumn}
\usepackage{bm}
\usepackage{amsfonts}
\usepackage{color}

\usepackage[normalem]{ulem}

\usepackage{multirow,array}

 \usepackage{url}
 
\begin{document}

\title{Cooperation among tumor cell subpopulations leads to intratumor heterogeneity}
\author[1]{Xin Li\thanks{xinlee0@gmail.com}}
\author[1]{D. Thirumalai\thanks{dave.thirumalai@gmail.com}}
 \affil[1]{Department of Chemistry, University of Texas at Austin, Texas 78712, USA}

\date{\today}

\maketitle

\begin{abstract}
\vspace{3mm}

\noindent  Heterogeneity is a hallmark of all cancers.  Tumor heterogeneity is found at different levels -- interpatient, intrapatient, and intratumor heterogeneity. All of them pose challenges for clinical treatments. The latter two scenarios can also increase the risk of developing drug resistance. Although the existence of tumor heterogeneity has been known for two centuries, a clear understanding of its origin  is still elusive, especially at the level of intratumor heterogeneity (ITH). The coexistence of different subpopulations within a single tumor has been shown to play crucial roles during all stages of carcinogenesis. Here, using concepts from  evolutionary game theory and public goods game, often invoked in the context of the tragedy of commons, we explore how the interactions among subclone populations influence the establishment of ITH. By using an evolutionary model, which unifies several experimental results in distinct cancer types, we develop quantitative theoretical models for explaining data from {\it in vitro} experiments involving pancreatic cancer as well as {\it vivo} data in glioblastoma multiforme.   Such physical and mathematical models  complement experimental studies,  and could optimistically  also provide new ideas for  the design of efficacious therapies for cancer patients.

\end{abstract}

\vspace{3mm}


\newpage

\section{Introduction}

Cancer is frequently described as a genetic disease arising through clonal evolution of cells. It is well appreciated that cancer is not just one but a group of diseases. Based on the original cell types and organs,  cancer has been classified into more than one hundred different types\cite{hanahan2000hallmarks}. From the latest statistics in the United States (see Table\ref{table1}), it is estimated that there are more than 1.5 million people are diagnosed as cancer patients, and over one third of them would die in 2019\cite{siegel2019cancer}.  Cancer is the second leading cause of death worldwide, and it also leads to the highest economic loss, which is estimated at approximately  $\$1.16$ trillion in 2010\cite{wang2016global,steward2014world}. Therefore, cancer has become not only a major public health issue but also  imposes a great economic burden on the society.

However, we still do not have any effective cure or even the ability to control most cancer types, although significant breakthroughs have been achieved in the past few decades for cancer prevention, understanding and even treatment in certain cases\cite{hanahan2000hallmarks,heymach2018clinical}. The difficulty in the war against cancer  is due to the complexity of this fatal disease, which exists in many different forms. Furthermore, the highly variable evolutionary properties also makes it extremely difficult to treat\cite{hanahan2000hallmarks,Nowell1976}.  In addition, cancer is a massively heterogeneous disease at several different levels\cite{Bedard2013,fox2014cancer}. Even within the same tumor, it can contain many different subpopulations with either genetic or epigenetic variations\cite{Gerlinger2012}. Another aspect which makes the disease even more difficult to study is that cancer cells are not independent moieties but rather should be viewed as an evolving ecosystem\cite{tabassum2015tumorigenesis}. Each cancer cell competes with others for limited resources and space subject to Darwinian evolution\cite{Nowell1976,greaves2012clonal}. Also it is frequently found that different cancer cell subpopulations cooperate with each other to overcome many biological constraints during their development\cite{tabassum2015tumorigenesis}. Therefore, we might need new methods and models  to understand the complexity of cancer at single cell level ($\mu m$ size) to tumor (several milli meters), and finally at the scale of an individual (meters).

In this article, we first briefly review the complexity of cancer and discuss some progress in modeling cancer progression with particular emphasis on intratumor heterogeneity (ITH). Then, illustrate the use of the mathematical model  to  explain the phenomena of ITH observed {\it in vitro} experiments on pancreatic cancer and an {\it in vivo} study on glioblastoma multiforme (GBM). The theoretical study also provides some potential insights into cancer treatment methods.

\section{Complexity of Cancer}
\subsection{Cancer types}
A recent compilation shows that there are about forty types of cancer based on the site of origin, such as lung, kidney, colon cancers and so on (see Table~\ref{table1})\cite{siegel2019cancer}. Due to the coexistence of different cell types in some organs,  cancers can be further classified as carcinoma (epithelial cells), sarcoma (connective tissue),  myeloma (plasma cells), leukemia (bone marrow), and lymphoma (cells of the lymph system), blastoma (precursor cells) etc. If we take kidney cancer as an example, there are renal cell carcinoma, transitional cell carcinoma, nephroblastomas, and renal sarcoma. The renal cell carcinoma (RCC) is the most common kidney cancer,  contributing to 90 percent of the cases, and it can be classified into a few subgroups (clear cell RCC, Papillary RCC, Chromophobe RCC, duct RCC and others) according to the phenotype of cancer cells\cite{ferlay2001globocan,cheville2003comparisons}. Indeed, it is quite astonishing that there are more than 100 different types of cancer known at present (see the full cancer list in $https://www.cancer.gov/types\#k$)\cite{hanahan2000hallmarks,canceratoz}, which explains the difficulty in determining broad principles that drive the origin and evolution of these diseases.  Yes, we are facing not a single disease but hundreds of them. Although all of these diseases share some similar hallmarks as summarized in the landmark reviews\cite{hanahan2000hallmarks,hanahan2011hallmarks}, there appears to be no universal treatment (nor will there be in the near future) for these diseases due to the lack of understanding of the underlying mechanisms of cancer evolution.

\subsection{Evolution of cancer} 
In addition to the many faces of cancer that  appear in different organs, another challenge for cancer treatment comes from tumor evolution. As proposed by Nowell in 1976, cancer is now widely viewed as a clonal evolutionary process\cite{Nowell1976,greaves2012clonal}. Due to the imperfect genome replication process, environment exposures and heredity, a normal cell can transform into a cancer cell by  acquiring genetic mutations, which occur on the time scale of a few decades\cite{vogelstein2013cancer,greaves2012clonal}. During cancer progression, tumor cells constantly face the selective pressures derived from their complex microenvironment, such as competition from the surrounding healthy cells for nutrients and space for growth\cite{moreno2008cell}. In addition, hypoxia, pH change, immune surveillance and other potential factors all threaten the survival and growth of tumor cells\cite{ackerman2014hypoxia,liotta2001microenvironment,balkwill2012tumor}.   Therefore,  tumor cells generate new traits through continuous evolution,  which is just one of the major reasons for the failure of chemotherapy, radiotherapy and other widely applied methods used to treat cancer patients\cite{Nowell1976,greaves2012clonal}. 

\begin{figure}
\vspace*{-0.1cm}
\hspace*{-0.2cm}
\centering
\includegraphics[clip,width=1.0\textwidth]{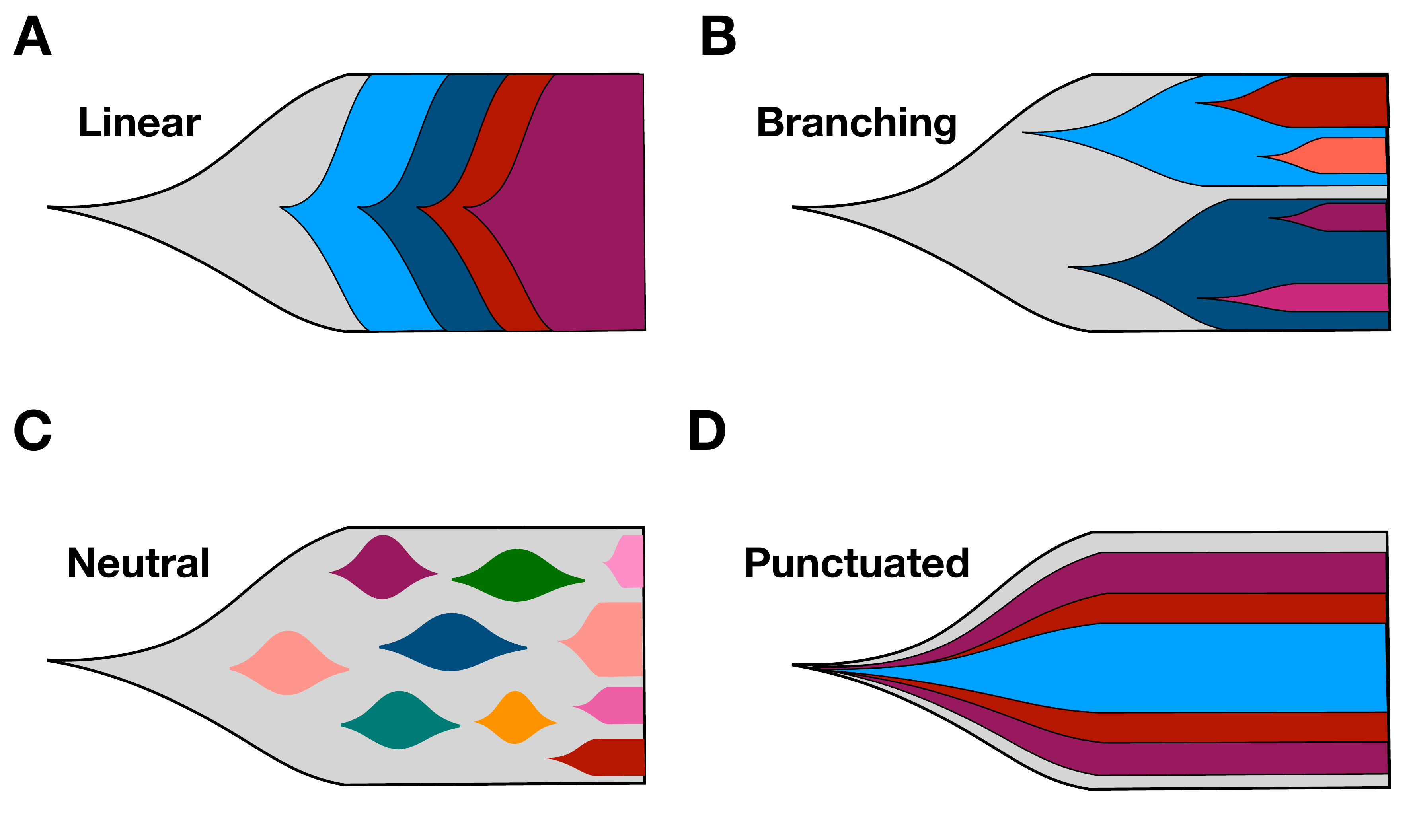}
\caption{\label{fig:linearbranch} {\bf  Four tumor evolution models.}  (A) Linear evolution: Cells accumulate driver mutations (with selective advantage for the cancer cells) sequentially during which selective sweeps occur. (B) Branching evolution:  New mutations appear before selective sweeps are completed and divergent subclones emerge from the same ancestor. (C) Neutral evolution: Only passenger mutations (without any selective advantage) accumulate stochastically in tumor cells. (D) Punctuated evolution:  Most of the detectable subclonal mutations occur  in short bursts of time at early stages of cancer evolution.  Different colors represent subclones with different mutations.}
\end{figure}

We have gained considerable knowledge from studies of tumor evolution through animal models\cite{van2002cancer,frese2007maximizing}.  The advent of many modern techniques, such as next generation sequencing have helped us get a deeper understanding for the genetic basis of cancer\cite{Vogelstein13}. However, it is still very difficult to study  tumor evolution in humans due to the inaccessibility of tumor biopsies from patients at different time points. Instead, the tumor evolutionary history obtained in most studies are derived from patient samples at a single time point based on assumptions, which are frequently violated as the tumor evolves\cite{davis2017tumor}. Hence, a clear and complete picture about tumor evolution has not emerged.  Many  models have been proposed to describe the tumor evolution process, such as the linear sequential model, branched, neutral and punctuated evolution of tumors (see Fig.~\ref{fig:linearbranch})\cite{davis2017tumor,gupta2017intratumor}. One common feature in all these models is that different subpopulations could appear and coexist in a tumor, which is the cause of pervasive cancer heterogeneity. 

\subsection{Cancer heterogeneities} 
\begin{figure}
\vspace*{-0.1cm}
\hspace*{-0.2cm}
\centering
\includegraphics[clip,width=1.0\textwidth]{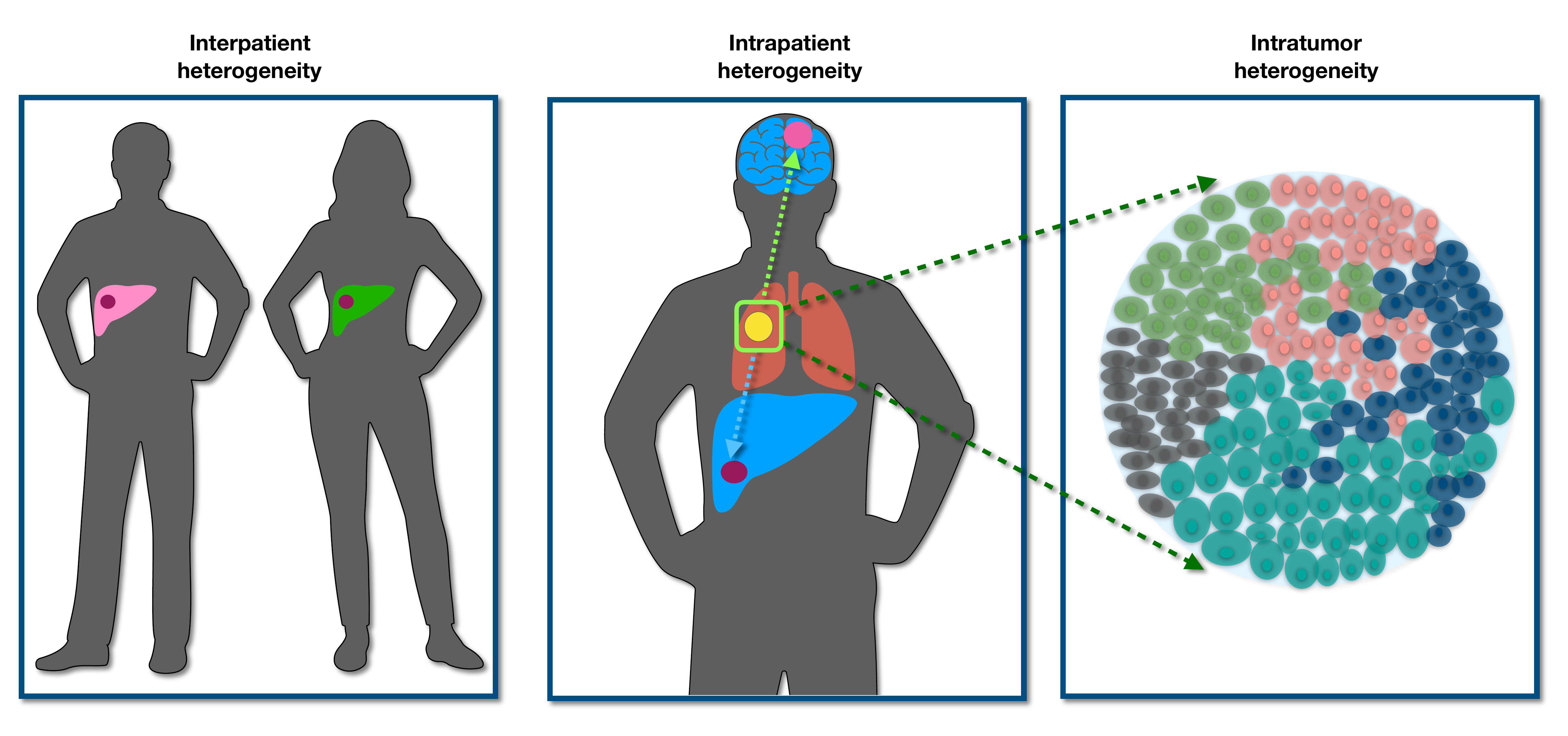}
\caption{\label{fig:levelofheterogeneity} {\bf Cancer heterogeneity.} Three levels of cancer heterogeneity: interpatient, intrapatient, and intratumor heterogeneity. Different color dots represent varied tumors or tumor cells with different genetic mutations. }
\end{figure}

Cancer is driven by accumulation of genetic mutations especially the ``driver mutations", which convey selective advantage to cancer cells.  With the advent of sequencing technology,  genome-wide association studies (GWAS) is affordable, thus giving us a powerful tool to search for cancer driver genes. The  list of cancer driver genes is continuously growing and the number has reached around 300 recently\cite{futreal2004census,bailey2018comprehensive}. A few cancer genes such as TP53 appear in many cancers while each cancer type usually has its own specific driver mutations. Therefore, different cancer patients have distinct tumor evolutionary processes, which leads to the interpatient heterogeneity (see Fig.~\ref{fig:levelofheterogeneity})\cite{Bedard2013,fox2014cancer}.  Personalized cancer medicine has been proposed, and is necessary due to this type of heterogeneity\cite{chin2011cancer}. 

As the cancer cells escape the primary site and seed  other sites of the body, they finish the transition into the last and fatal stage, referred to as cancer metastasis - responsible for $90\%$ of cancer patient death\cite{chaffer2011perspective}. Whether new driver mutations are required for the cancer metastasis is unresolved\cite{robinson2017integrative,yates2017genomic,priestley2019pan,birkbak2020cancer}. However, distinct new mutations can appear both at the primary and metastatic sites after the spatial isolation is established among cells at different sites\cite{turajlic2016metastasis}. It leads to the next level of heterogeneity, intrapatient heterogeneity which is one of the reasons for cancer recurrence after treatment (see Fig.~\ref{fig:levelofheterogeneity}).

In addition to the two types of cancer heterogeneities described above, the intratumor heterogeneity (ITH, see Fig.~\ref{fig:levelofheterogeneity}), which refers to the coexistence of different subpopulations in a single tumor has been found in many cancers\cite{Gerlinger2012,Sottoriva2013,Bashashati2013,Gerlinger20142,Bruin2014,Yates2015,Boutros2015,Ling2015}. ITH plays a crucial role in almost all stages of cancer such as tumor progression, metastasis, drug resistance and recurrence of cancer\cite{burrell2013causes,lawson2018tumour,janiszewska2019subclonal}. Therefore, understanding the ITH is a major step in investigating cancer heterogeneity because it can help design better therapy to avoid drug resistance and cancer recurrence. 

The evolutionary models mentioned above  all point to the possibility that different subpopulations (with distinct mutations) can coexist in a single tumor.  In these models, the subpopulations interact with each other mainly through cell competition irrespective of the selective action under linear and branched models or the drift effect in neutral and punctuated models. However, increasing experimental evidence found that the cooperation among  distinct cell subpopulations in a tumor  is essential for tumor maintenance~\cite{Cleary14}, enhanced tumor growth~\cite{Marusyk2014}, and even cancer metastasis~\cite{janiszewska2019subclonal,Chapman2014,Aceto2014}. Surprisingly, a minor subpopulation is sufficient to support the whole tumor growth and determine the clinical course~\cite{Mullighan2008,Johnson2014,Morrissy2016}. Hence, the different types of cell-cell interactions have to be considered to better understand the mechanisms of ITH, which is often neglected in many theoretical models. The lack of understanding of ITH greatly impairs the progress of developing more effective therapies for cancer patients.

\section{Mathematical models of ITH} 
\subsection{The multistage model of cancer evolution} 
In the past few decades, many innovative mathematical models have been proposed to study the cancer evolutionary process, including cancer initiation, progression, and metastases\cite{altrock2015mathematics}. A simple multistage model that cancer is driven by a number of driver mutation events was proposed to explain the age-dependent cancer incidence rate more than sixty years ago\cite{armitage1954age}.  The Moran process accounts for the cell division, apoptosis, and mutation process in  tumor evolution  for a fixed cell population size. It has been widely used to investigate the accumulation of mutations and cancer initiation\cite{michor2004dynamics,beerenwinkel2007genetic,merlo2006cancer,tomasetti2013half,teimouri2019elucidating}. Other models which further include the spatial structure of tumors and its related microenvironment such as the oxygen and nutrient concentrations are considered to investigate the growth dynamics of tumor by using partial differential equations or an agent-based model\cite{sherratt2001new,nowak2003linear,beerenwinkel2014cancer,malmi2018cell,sinha2019spatially}.

\subsection{ITH under competition and cooperation} 
Evolutionary models  have also been used to study ITH\cite{kansal2000emergence,bozic2010accumulation,durrett2011intratumor,waclaw2015spatial}, and many insightful results have been obtained from these studies, such as the extent of ITH and factors influencing ITH.  There are two crucial assumptions in  these theoretical studies.  The genetic mutations only confer fitness advantage to the cell itself (cell-autonomous effect), and tumor cells interact with each other through competition.  However, there is increasing evidence   that  tumors cannot overcome the microenvironmental constraints only through autonomous increase of the cell growth rate\cite{gatenby2008microenvironmental,bissell2011don,quail2013microenvironmental,curtius2018evolutionary}. In contrast, enhancement of survival and proliferation rates through non-cell-autonomous effects by factors such as  metalloproteinases and cytokines are critical for tumor progression\cite{marusyk2014non}. Therefore, these factors secreted by certain tumor/normal cells (`producer') can bring cooperation among different cell subpopulations instead of competition. Such a cooperation can in principle accelerate  tumor progression.  Instead of one population accumulating all the mutations required for cancer development\cite{hanahan2011hallmarks},  a few  partially transformed subpopulations with each containing one or two mutations can realize this procedure through their cooperation\cite{tabassum2015tumorigenesis,axelrod2006evolution}.

\subsection{Evolutionary game theory} 
The evolutionary game theory (EGT) provides a novel and unique avenue for investigating ITH accounting for the interactions among subpopulaitons of tumor or between cancer and normal cells\cite{dingli2009cancer,basanta2012investigating,li2019share,archetti2019cooperation,gluzman2018optimizing,stavnkova2019optimizing}.  The EGT is a subfield of game theory (GT), which provides mathematical models for studying the strategic interaction among individuals\cite{smith1982evolution,osborne2004introduction}. An individual (`player') receives a payoff during the game depending on both the player and also the  behavior (`strategy') of others. For EGT in cancer research, the players are cancer or normal cells, and the payoff is their fitness while the strategies are phenotypes adopted by players\cite{archetti2019cooperation}.  The advantage of EGT is that it can describe the time-dependent evolution of the relative abundance of each cell type, determine the equilibrium conditions and the stability of phenotype coexist\cite{altrock2015mathematics}.  Here, we briefly describe the EGT. The evolutionary games for two cells of different types $A$, and $B$ are frequently represented by the pay-off matrix,
  \begin{table}[h]
    \setlength{\extrarowheight}{2pt}
    \begin{tabular}{ccc|c}
      & \multicolumn{1}{c}{} & \multicolumn{1}{c}{$A$}  & \multicolumn{1}{c}{$B$} \\\
     \multirow{2}*{}& $A$ & $W_{AA}$ & $W_{AB}$ \\\cline{3-4}
      & $B$ & $W_{BA}$ & $W_{BB}$ \\\
    \end{tabular}
  \end{table} \\
where $W_{IJ}(I,J\equiv A, B)$ represents the fitness of the cell type $I$  interacting with the cell type $J$. The cell-cell interaction can be direct or indirect and its effect can also be  competition or cooperation due to the influences of space, nutrients, information, growth factors and  other microenvironmental factors\cite{moreno2008cell,archetti2019cooperation,pacheco2014ecology}. Therefore, the fitness function $W_{IJ}$ of the cells can also be expressed by very different mathematical formula\cite{archetti2019cooperation}.  As the two different cell populations, instead of just two cells, are well-mixed with each other\cite{pacheco2014ecology}, the average fitness ($w_{A}$, $w_{B}$) of the two cell types may be described as,
  \begin{equation}
\label{fit-ex}
w_{A} =  f_{A}W_{AA}+(1-f_{A})W_{AB}\, ,
\end{equation} 
\begin{equation}
\label{fit-ex}
w_{B} =  f_{B}W_{BA}+(1-f_{B})W_{BB}\, ,
\end{equation} 
 where $f_{A}$, and  $f_{B}$ are the fractions of the cell type $A$, and $B$ in the population, respectively. From the fitness functions of the cells, the time-(in)dependent properties of the cell population as mentioned above can be calculated. 
 
In a few latest studies\cite{Archetti15,kimmel2019neighborhood}, the EGT has been adopted to explain ITH in pancreatic cancer starting from the simplest case with only two different types of cancer cells. One cell type can produce a growth factor while the other does not.  The growth factor (often called ``public goods") can be consumed by both cell types and promote their proliferation. Qualitative conclusions are obtained for the coexistence of two cell subpopulations which are consistent with experimental observations.

\section{Public goods game}
In a recent study\cite{li2019share}, we  investigated the ITH through the ``public goods game" among cancer cells quantitatively. Instead of assuming a constant population size, and ideal fitness functions, we consider a growing cell population, and took cell growth rate as the fitness function. Both these quantities can be measured from experiments directly.  In the following, we will briefly introduce the model and discuss some of the most important results that we discovered. 

The public goods game is an economic model, which has  extensive applications in many  areas, such as microbial colonies, and insect communities, and even cancer research~\cite{Hofbauer1998,Hauert2006,nowak2006five,Allen2013,Nanda2017,bauer2018role,archetti2019cooperation}.  There are two players in this model based on whether they produce the public goods (called `producer') or not (`non-producer'). Both of them derive benefits from the public goods while the producer usually has to pay a cost for the public goods production. We investigate the underlying mechanism of ITH during cancer progression within the framework of public goods game.  We will give two examples to show how cooperation among cancer cell subpopulations can lead to the establishment of a stable ITH, which could be applicable to many other cancers.

\subsection{Models}
\begin{figure}
\vspace*{-0.1cm}
\hspace*{-0.2cm}
\centering
\includegraphics[clip,width=1.0\textwidth]{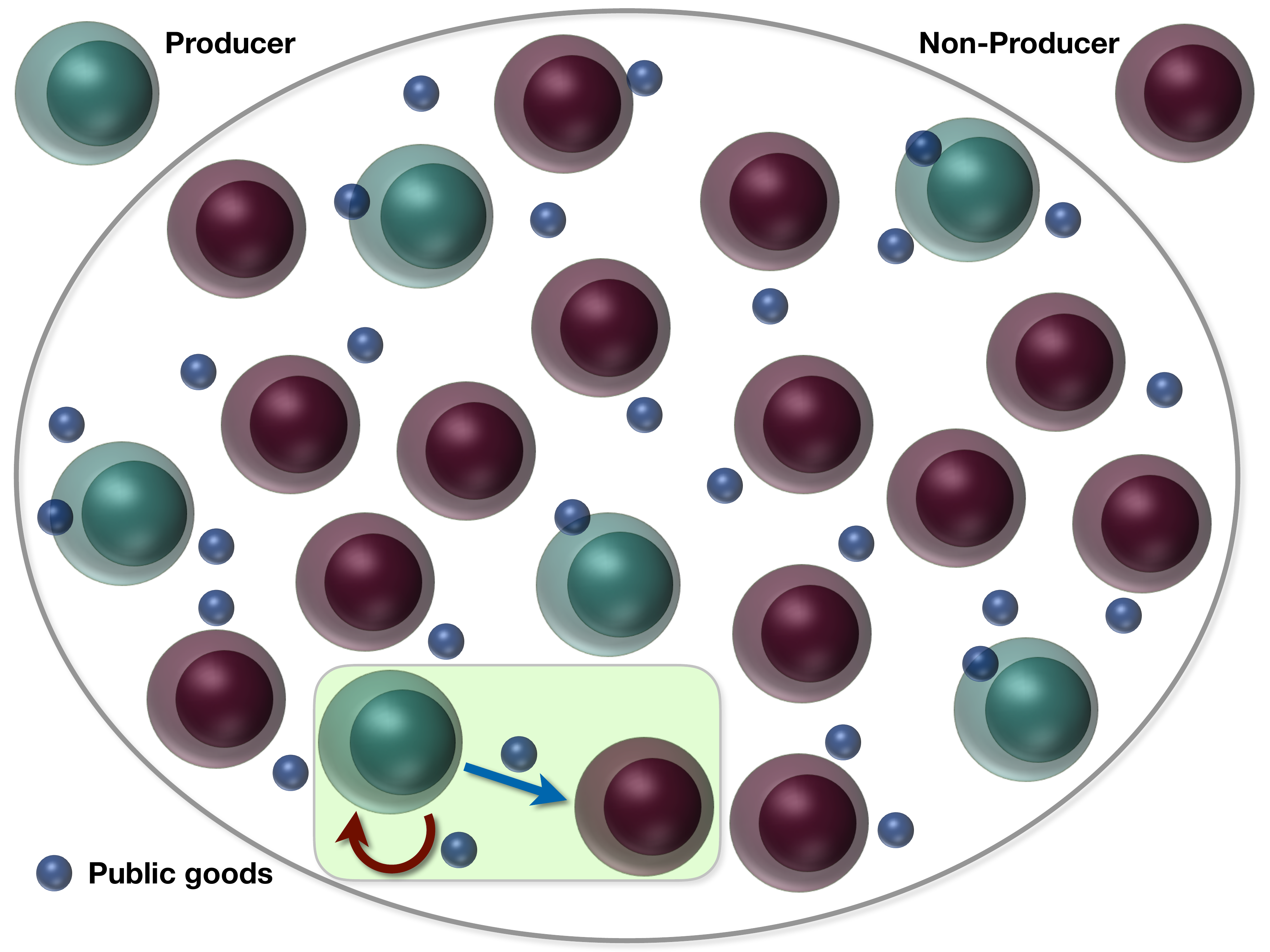}
\caption{\label{fig:schematicfig} {\bf Schematic figure for the public goods game of two cell types.}  The producers (cyan cells) generate public goods (small green dots) which are shared among producer and non-producer cells (magenta cells). The public goods promote the proliferation of both cell types. The inset rectangle: A detailed picture shows the circuit of public goods production and the feedback for the proliferation of producer and non-producer. }
\end{figure}

We consider two types of cancer cells with one of them producing a public good (see Fig.~\ref{fig:schematicfig}). In the insightful {\it in vitro} experiment\cite{Archetti15},  the public good is the insulin-like growth factor II (IGF-II). First, we focus on this experiment and then we apply the same idea to describe the ITH observed in an {\it in vivo} experiment on glioblastoma multiforme\cite{Inda2010}. In the first experiment, both cell types are taken from cancer cells of mice with insulinomas (a type of pancreatic cancer).   One cell type can produce IGF-II (the producer (+/+)) while the non-producer (-/-) cells have IGF-II gene deleted, which means they can no longer produce IGF-II but can still consume it from the environment. 

One critical element in using the concept of the public goods game is how to define the payoff function. Often a rather complex, frequency dependent function is assigned to both players\cite{Hauert2006,Archetti15}.  Here, we use the growth rate as the fitness functions for both the cell types  as they can be measured in experiments directly.  A non-linear growth function\cite{Archetti15} is found for the growth rate ($w_{-}$) of the non-producer cells as a function of  IGF-II concentration ($c$), which is well-described by the Hill-like function,
\begin{equation}
\label{fit-ex}
w_{-} = a_{1} + \lambda_{1}c^{\alpha}/(a_{2}^{\alpha}+c^{\alpha}) \, .
\end{equation} 
where $a_{1} = 2.0$, $\lambda_{1} = 18.9$, $\alpha = 0.7$, and $a_{2} = 3.2$\cite{li2019share}. Although it has not been measured, we expect that the producer cells  also have a similar functional form as Eq.~(\ref{fit-ex}) for their fitness ($w_{+}$), which leads to,
\begin{equation}
\label{fit+ex} 
w_{+} = g(c) -p_{0} \, ,
\end{equation} 
where $g(c)$ is the same Hill-like function as in Eq.~(\ref{fit-ex}), and $p_{0}$ is the cost paid by producer cells due to the IGF-II production.  The IGF-II concentration ($c$) has two sources once two cell types are mixed with each other. One is from the production of the producer cells, and the other is supplied exogenously (in the culture medium). The IGF-II produced by the producer will depend on the fraction $f_{+}$ in the population. Therefore, the IGF-II concentration ($c_{+}$) available for the producer can be written as, 
\begin{equation}
\label{cplus} 
c_{+} = af_{+}+c_{0} \, ,
\end{equation} 
where $a$ is the coefficient for the allocation of IGF-II produced by +/+ cells,  and $c_{0}$ represents the exogenous supply of IGF-II.    Similarly, the available IGF-II ($c_{-}$) for  the non-producer is given by,
\begin{equation}
\label{IGFIIC}
 c_{-}  = b f_{+} + c_{0} \, , 
\end{equation} 
The parameter $b$ is the coefficient of allocation of IGF-II produced by +/+ cells. We use different parameters $a, b$ to indicate that the producer and non-producer can get different fraction of IGF-II that are produced by the former cells.

The evolution of the fraction $f_{+}$ ($f_{-}$) of producers (non-producers) can be derived from the replicator equations,
\begin{equation}
\label{f+}
\frac{\partial f_{+}}{\partial t} = (w_{+}-\langle w \rangle)f_{+}\, , 
\end{equation} 
and,
\begin{equation}
\label{f-}
\frac{\partial f_{-}}{\partial t} = (w_{-}-\langle w \rangle)f_{-}\, ,  
\end{equation} 
where the average fitness $\langle w \rangle$ is,
\begin{equation}
\label{fitaverage}
\langle w \rangle = w_{+} f_{+} + w_{-}f_{-}\, , 
\end{equation} 
with $f_{+} + f_{-} = 1$ being the normalization condition. Let the number of producers and non-producers be $N_{+}$, and $N_{-}$, respectively. The whole population size, $N$,  is the sum of $N_{+}$ and $N_{-}$. The system size, $N$, is a time dependent quantity, which  is often neglected in other models for simplicity~\cite{Archetti15,Blythe07}. Here, we describe the time-dependent changes in the system size, $N$, through
\begin{equation}
\label{Number}
 \frac{\partial N}{\partial t} =w_{+}N_{+}+w_{-}N_{-} = \langle w \rangle N \, . 
\end{equation} 
Then, we can study the conditions for the coexistence of producers and non-producers from the equations above.

\subsection{Unequal allocation of public goods results in stable ITH}
\begin{figure}
\vspace*{-0.9cm}
\hspace*{-0.2cm}
\centering
\includegraphics[clip,width=0.75\textwidth]{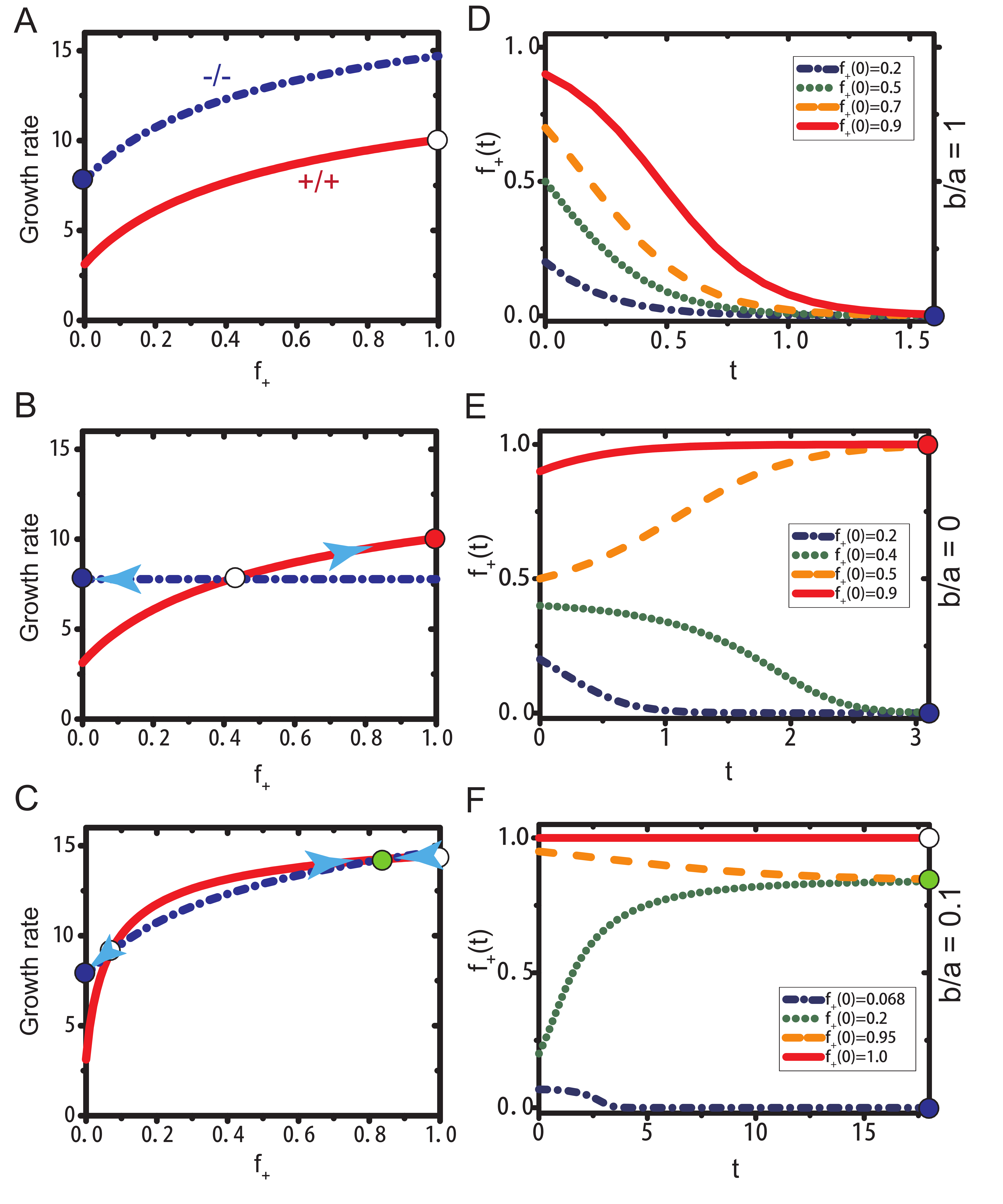}
\caption{\label{fig:Share} (A)-(C): The growth rates of producer (+/+) and non-producer (-/-) cells as a function of the +/+ cell fraction ($f_{+}$) at different allocation strategies of IGF-II generated by the +/+ cells. (A) The IGF-II are  shared equally between two cell types ($b=a=8$), (B) +/+ cells do not share any IGF-II with -/- cells ($b=0$, and $a=8$), (C) +/+ cells only share a small fraction of IGF-II with -/- cells  ($b=8$, and $a=80$). The parameter value $c_{0} =1$ and $p_{0}= 4.65$. The growth rate of +/+ (-/-) cells are shown in solid red (dash-dotted blue) lines. The stable or unstable fixed points are represented by filled or empty circles, respectively. (D)-(F):  The fraction $f_{+}(t)$ of +/+ cells as a function of time at different initial conditions corresponding to the figures (A)-(C) on the left, accordingly.   The growth rate is derived from the cell's relative density change at the log phase~\cite{Archetti15}. The time  unit is in days. Figure is adopted from Ref.~\cite{li2019share}.}
\end{figure}

The first important result derived using our model is that the emergence of a stable ITH requires an unequal allocation of public goods between producer and non-producer cells.  The allocation of the public goods produced by the producers is determined by the ratio, $b/a$, in Eqs.~(\ref{cplus}) and (\ref{IGFIIC}).  A unity for the value of $b/a$ means that the two cell types share the public goods equally. Therefore, the two cell types will have the same fitness function, expect a constant shift $p_{0}$ due to the cost paid by the producer (see Eqs.~(\ref{fit-ex}) and (\ref{fit+ex}) and the Fig.~\ref{fig:Share}A). There is only one stable state under this condition (see the filled blue dot in  Figs.~\ref{fig:Share}A and D),  a homogeneous state consisted of only non-producer would be expected as long as the initial fraction $f_{+}(0)<1$.  On the other hand, given $b/a = 0$ which means producers do not share any public good with the non-producer. Then, the fitness of the non-producer does not depend on the fraction of the producer any more. There is a heterogeneous state but it is unstable under this condition (see the open circle in  Fig.~\ref{fig:Share}A and the evolution of $f_{+}$ in  Fig.~\ref{fig:Share}E). However, we find a stable heterogeneous state if $0< b/a <1$ (see the filled blue dots in Fig.~\ref{fig:Share}C and F), which indicates that the producer maintains a higher fraction of the public goods produced. There is a negative feedback mechanism close to this stable coexisting state, which can be observed from Fig.~\ref{fig:Share}C. 

\begin{figure}
\vspace*{-0.1cm}
\hspace*{-0.2cm}
\centering
\includegraphics[clip,width=1.0\textwidth]{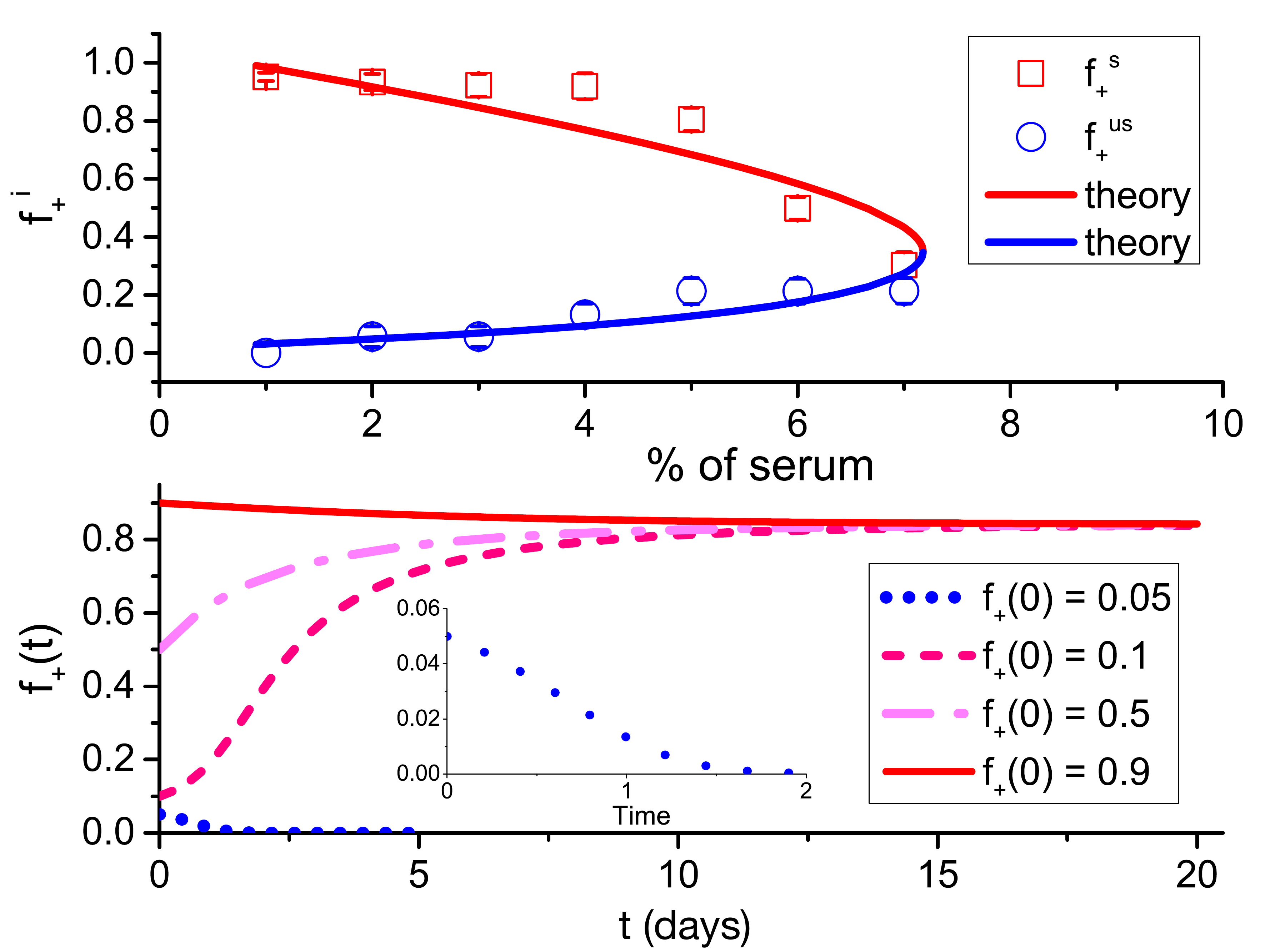}
\caption{\label{fig:phase} Upper panel: The producer fraction $f_{+}^{i}$,( $i \equiv s$ or $us$, with $s$ for stable and $us$ for unstable state) at internal equilibrium states (observed on day 5 under different initial fractions of $f_{+}(t=0)$ in experiments) as a function the levels of serum.  The red squares (blue circles) represent stable (unstable) states. The error bars indicate the upper and lower boundaries observed in experiments. Our theoretical predictions are illustrated by the solid lines. The parameter values:  $a = 80$, and $p_{0} = 4.65$. Lower panel:  The time dependence of  $f_{+}(t)$ (the fraction of +/+ cells) at $3\%$ of serum under various initial conditions.  The inset is a zoom in for $f_{+}(t)$ with $f_{+}(0) = 0.05$.   Figure is adopted from Ref.~\cite{li2019share}.}
\end{figure}

To validate our conclusion that it is the unequal allocation of public goods leading to the establishment of a stable coexist state, we calculated the the internal equilibrium fractions ($0 < f_{+}^{i} <1$) of +/+ cells at different concentrations of serum ($c_{0}$) which has been detected in experiments.  The +/+ cell fraction is measured after 5 days co-culture of +/+ and -/- cells under eleven different initial fractions of cells (with $f_{+}(t=0)$ almost evenly distributed between 0 and 1) and different amounts of serum\cite{Archetti15}. There are three free parameters in our model, a, b, and $p_{0}$.  From the equal fitness ($\approx$ 14.4) of producer and non-producer cells and the +/+ cell fraction approaching 1 at the stable internal state under $c_0 = 0$ observed in the experiments\cite{Archetti15}, the parameter b $\approx$ 8 is obtained from Eqs.~(\ref{fit-ex}) and (\ref{IGFIIC}). Similarly, we can derive the values of $a$ and $p_{0}$ based on the experimental results of internal equilibrium states. We find that our theoretical model captures the experimental observations very well (see the upper panel in Fig.~\ref{fig:phase}). No quantitative explanations have been reached for similar experiments from other models at present.   Interestingly, the ratio of $b/a$ obtained through the comparison between our theoretical results and experiments is 0.1 which is just located at the interval we proposed above for the formation of a stable ITH (see the low panel of Fig.~\ref{fig:phase} for the illustration of one stable ITH state).


\subsection{Cost $p_{0}$ influence cell cooperation}
\begin{figure}
\vspace*{-0.1cm}
\hspace*{-0.2cm}
\centering
\includegraphics[clip,width=1.05\textwidth]{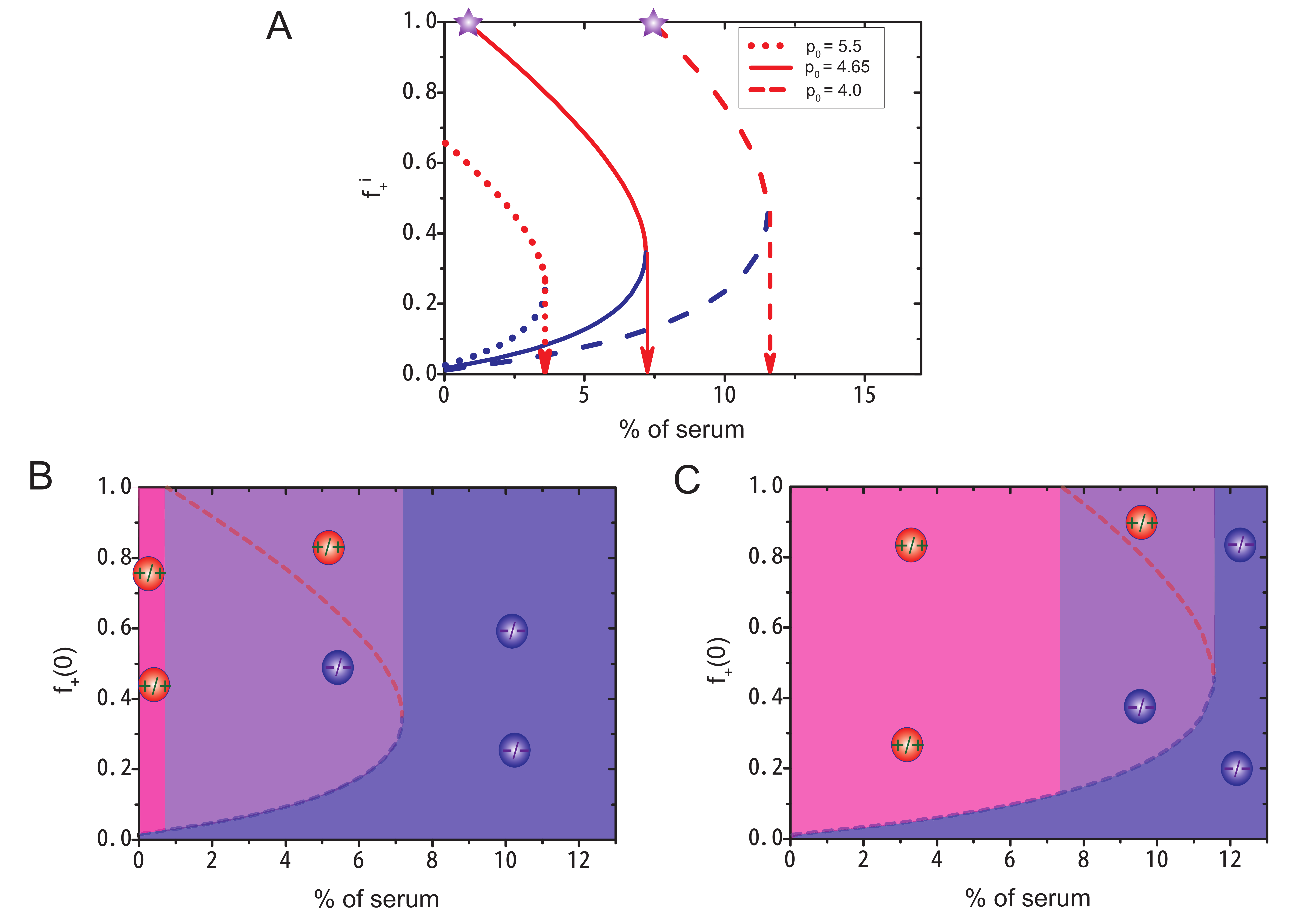}
\caption{\label{fig:price} (A) The producer fraction $f_{+}^{i}$, ($i \equiv s, us$) at internal equilibrium states as a function of the serum level at different values of $p_{0}$. The red (blue) lines indicate stable 
(unstable) equilibrium fractions $f_{+}$. Arrows and purple stars indicate the two critical serum concentrations. (B) and (C) Phase diagrams in terms of the initial fraction $f_{+}(0)$ and  \% of serum. The values of $p_{0}$ in (B) and (C) are 4.65 and  4.00, respectively. There are three stable phases in these two figures. (i) A homogeneous phase with producers only (see the region in pink color). (ii) A heterogeneous phase with both producers and non-producers (purple color). The red dashed line indicates the stable equilibrium fraction of producers. (iii) A homogeneous phase with non-producers only (see the blue color). The red and blue spheres labeled by +/+ and -/- represent the producer and non-producer cells, respectively. Figure is adopted from Ref.~\cite{li2019share}.}
\end{figure}

Another important result is that the cost $p_{0}$ paid by producers has a strong influence on cell cooperation. As the fitness of producers is reduced by $p_{0}$ due to the production of the public goods, we investigate whether this parameter has any influence on the cooperation of different subpopulations. From the diagram of the internal equilibrium fraction $f_{+}^{i}$ of the  producer as a function of the exogenous resource level at varied values of $p_{0}$ (see Fig.~\ref{fig:price}),  we found a critical concentration for exogenous resources above which there is only one stable homogeneous state consisted of non-producers.  And this critical concentration increases as the value of $p_{0}$ decreases. A second critical value (see the star symbols in Fig.~\ref{fig:price}A) is observed for exogenous resources as $p_{0}$ takes small values. There is no stable heterogeneous states after the exogenous resource level is lower than this critical value.  

To better understand the influence of $p_{0}$ on cell cooperation, we calculated a phase diagram in terms of the initial fraction $f_{+}(0)$ and percentage of serum (see Figs.~\ref{fig:price}B and \ref{fig:price}C). There are three distinct phases represented in different colors (pink, blue, and purple). Two stable homogeneous states consist of either producers (pink region) or non-producers (blue region). The third phase corresponds to a stable heterogeneous state consisted of both subpopulations (purple region). The area of the three regions is strongly influenced by the value of $p_{0}$ (see Figs.~\ref{fig:price}B and \ref{fig:price}C). A heterogeneous phase is established at modest levels of exogenous resources and higher $f_{+}(0)$. We also find that a higher cost paid by the producers actually  helps the establishment of a heterogeneous system (see the change of the purple region in Figs.~\ref{fig:price}B and \ref{fig:price}C).

\subsection{Cooperation among cancer subpopulations in Glioblastoma multiforme} 
Cooperation among different cell types has been found in several types of cancer such as breast cancer\cite{Cleary14}, prostate cancer\cite{mateo2014sparc} and Glioblastoma multiforme (GBM)\cite{Inda2010}. Here, we take GBM, which is the most aggressive and fatal brain cancer~\cite{Gallego15} as an example, to show how the same model based on public goods explains equally well the mechanism of ITH in GBM. Two types of cancer cells are frequently found in GBM with different expressions of epidermal growth factor receptor (EGFR)~\cite{Nishikawa94}. Apparently, the coexistence of the two cell types in GBM promotes the cancer progression and leads to a worse prognosis of the patients~\cite{Shinojima03,Heimberger05}. Several studies found that the GBM cells with rearrangement of EGFR gene ($\Delta$ cells) can secrete factors (Interleukin-6 and Leukemia inhibitory factor) to enhance cell proliferation and inhibit its apoptosis~\cite{Inda2010,Heinrich03}.  

\begin{figure}
\vspace*{-0.1cm}
\hspace*{-0.2cm}
\centering
\includegraphics[clip,width=1.0\textwidth]{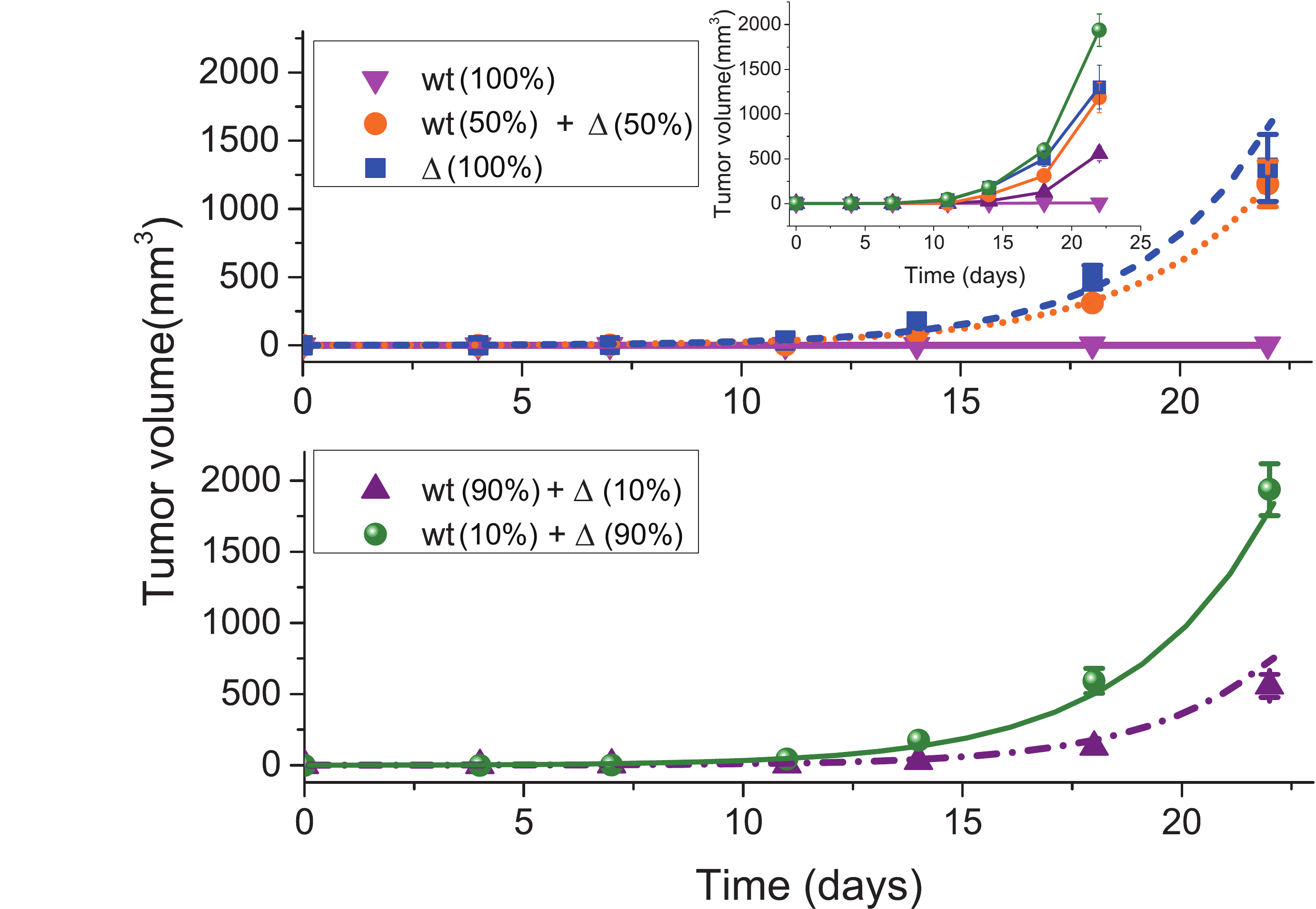}
\caption{\label{fig:tumorvolume} {\bf The evolution of the tumor volume.}  The GBM tumor volume as a function of time at different initial fractions of WT and  $\Delta$ cells.  The experimental data are represented by symbols. The  data from the upper panel are used to obtain the three free parameter values ($a = 68.4, b=0.946$ and $p_{0}=0.651$) in the model.  The purple and green curves at the lower panel are theoretical predictions. Error bars indicate the standard error. The inset gives the complete experimental data.  Figure is adopted from Ref.~\cite{li2019share}.}
\end{figure}

A recent study investigated the interactions between GBM cells with amplified levels of EGFR (referred to as  WT cells) and $\Delta$ cells systematically\cite{Inda2010}. A fixed amount of cancer cells ($2\times 10^5$ cells) with different initial fractions of the two cell types are injected into athymic nude mice (4 to 5 weeks old).  Then, the size of the established tumors was measured at different times (see the different symbols in Fig.~\ref{fig:tumorvolume}).    It is difficult for the WT cells alone to induce a new tumor in the mice (see the pink upside down-triangles in Fig.~\ref{fig:tumorvolume}) while a tumor can be quickly generated by the $\Delta$ cell alone (see the the blue squares in Fig.~\ref{fig:tumorvolume}). Meanwhile, a faster tumor growth rate is found as the initial fraction of $\Delta$ cells increases until it reaches 90$\%$ (see the inset of Fig.~\ref{fig:tumorvolume}). However, the maximum growth rate is found at the mixture of 10$\%$ WT and 90$\%$ $\Delta$ cells but not 100$\%$ $\Delta$ cells which seems to indicate a cooperative relation between these two cell types.

We applied the public goods model to explain the non-trivial experimental observations summarized above. The WT and $\Delta$ cells are considered to be the non-producer and producer, respectively. The IL-6 and/or LIF secreted by $\Delta$ cells is the public goods. We used a simple fitness function ($w_{-}$) for the WT cell as above which is given by, 
\begin{equation}
\label{fit-exsi}
w_{-} = b  f_{+}/(1+b f_{+}) \, .
\end{equation} 
The fitness function ($w_{+}$) for $\Delta$ cells is,
\begin{equation}
\label{fit+exsi}
w_{+} = a  f_{+}/(1+a f_{+}) -p_{0}\, .
\end{equation} 
The different parameters $a$ and $b$ indicate that the public goods (produced by the $\Delta$ cells) can be shared unequally between the two cell types. There are only three free parameters in our model, which can be calculated in the following way. 
 The average fitness of the population is given by $\langle w \rangle = w_{+}$ as it contains only producer cells with $f_{+} = 1$ (see Eq.~(\ref{fitaverage})). Therefore, the population size,  $N$, can be calculated from Eq.~(\ref{Number}) which increases exponentially, $N = N_{0}e^{w_{+}t}$. From the  growth curve (see the blue square in the upper panel of  Fig.~6) of the $\Delta$ cells measured in experiments, we obtain,
\begin{equation}
\label{fit+exsi2}
\frac{a} {1+a } -p_{0} \approx 0.335  \, ,
\end{equation} 
where 0.335 is obtained from the exponential fit. It is found that the two types of cells grow at the same rate  given $f_{+} = f_{-}$\cite{Inda2010}, which leads to,
\begin{equation}
\label{fit+exsi50}
\frac{0.5a} {1+0.5a } -p_{0} =\frac{0.5b} {1+0.5b}  \, .
\end{equation} 
In addition, the evolution of $N(t)$ can be described by  $N(t) = N_{0}e^{w_{-}t}$  given $f_{+} = f_{-}$, and $w_{+} = w_{-}$ as the average fitness $\langle w \rangle  = 0.5(w_{+}+w_{-}) =  w_{-}$. Thus, we obtain the following expression,
\begin{equation}
\label{fit+exsi501}
\frac{0.5b} {1+0.5b } \approx 0.321  \, , 
\end{equation}  
from the tumor growth curve (orange dots) in the upper panel of  Fig.~\ref{fig:tumorvolume}. The constant 0.321 is obtained from the exponential fit (see the orange dotted line in Fig.~\ref{fig:tumorvolume}). Therefore, the three parameters in our model can be derived using Eqs.~(\ref{fit+exsi2})-(\ref{fit+exsi501}), which leads to, $a = 68.4$, $b = 0.946$, and $p_{0} = 0.651$. 

To test our model, we predicted the tumor growth behavior at 10$\%$, and 90$\%$ of $\Delta$ cells, and found that our predictions agree very well with experimental observations (see the lower panel of Fig.~\ref{fig:tumorvolume}). We also calculated the tumor growth rate as a function of the fraction of $\Delta$ cells\cite{li2019share} and found that there is a maximum tumor growth rate at an intermediate value ($\approx$0.77) of the $\Delta$ cells fraction which is consistent with experimental observations and can be test in future experiments.

\begin{figure}
\vspace*{-0.1cm}
\hspace*{-0.2cm}
\centering
\includegraphics[clip,width=0.82\textwidth]{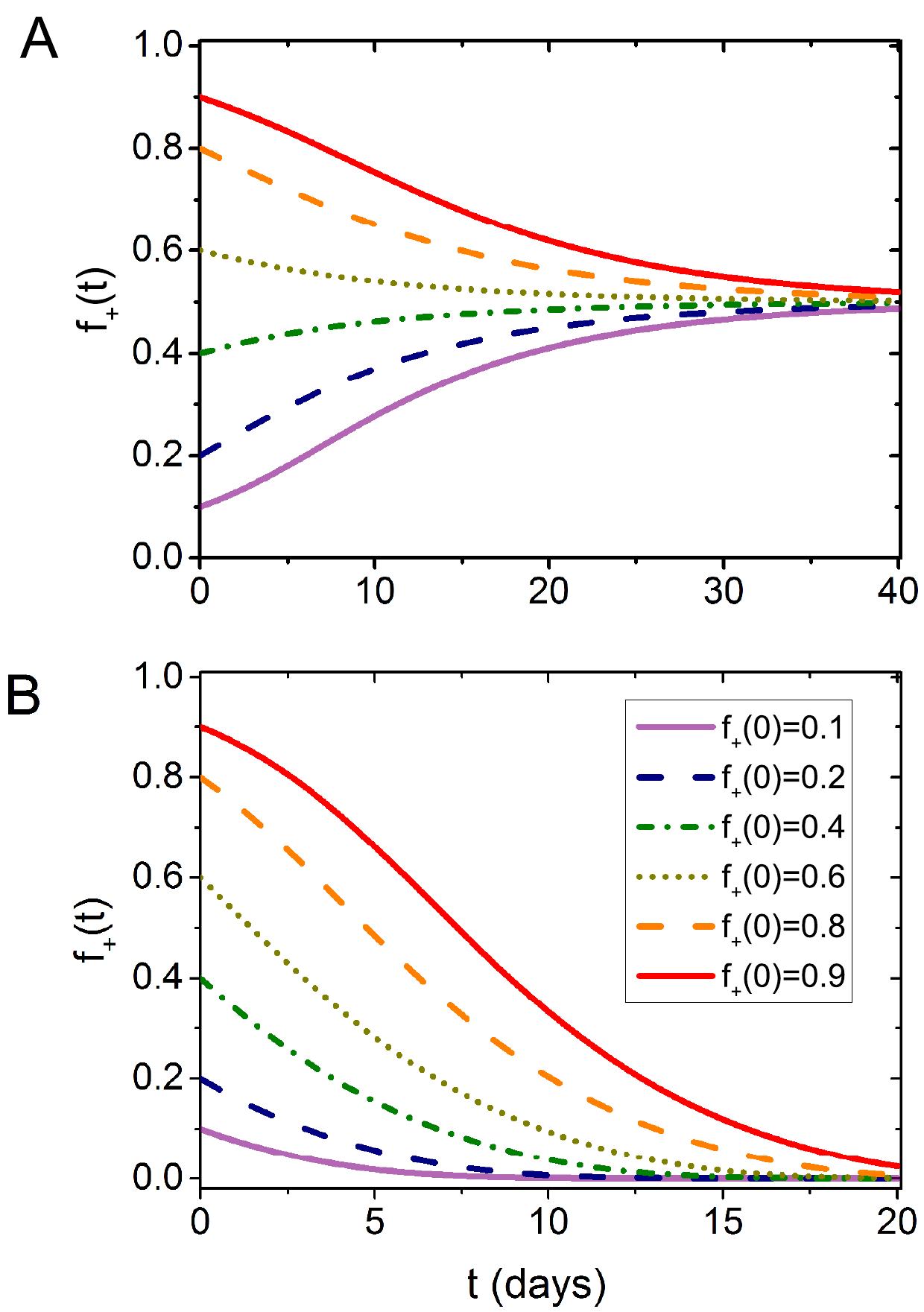}
\caption{\label{fig:effextrasource} Prediction of the fraction $f_{+}(t)$ of $\Delta$ cells as a function of time under distinct  initial conditions ($f_{+}(t=0)$) in GBM. (A) Without external supply of  public goods ($c_{0} = 0.0$);  (B) With external supply of  public goods, and the total public goods becomes $a  f_{+}+c_{0}$ ($b  f_{+}+c_{0}$) for the producer (non-producer) with $c_{0} = 1.0$. Other parameters are the same as in  Fig.~\ref{fig:tumorvolume}. The labels in (A) and (B) are the same. Figure is adopted from Ref.~\cite{li2019share}.}
\end{figure}

Our theory also provides some hints for the frequent observations of both WT cells and $\Delta$ cells in GBM patient and its poor prognosis. From the evolution of the fraction of $\Delta$ cells under different initial conditions (Fig.~\ref{fig:effextrasource}A), we found that the heterogeneous state is  stable even $f_{+}(0)$ varies from 0.1 to 0.9.  Therefore, a quick recurrence of GBM is possible if some $\Delta$ cells still exist in the tumor.  Cooperation between different cancer types greatly enhanced their survival and progression. Hence, new drugs and methods should be developed to eliminate such a relation. Our theory predicts that a rich culture medium can encourage cell competition instead of cooperation, which might provide potential treatment strategies for GBM and other similar cancer types (see Fig.~\ref{fig:effextrasource}B).

\section{Discussion}

In this article, we briefly reviewed different aspects of cancer complexity from the surprisingly  long list of cancers. How to deal with such a complex evolving system poses a big challenge for biologists, clinicians and presents new opportunities for physicists.  The ten hallmarks of cancer summarized by Hanahan and Weinberg in 2011 provide a guideline for understanding the progression of neoplastic disease\cite{hanahan2011hallmarks}. Due to the ubiquitous heterogeneity present in cancer, there is  no universal treatment method for all cancers although the latest immunotherapy brings great hope in this direction\cite{mellman2011cancer,ribas2018cancer}.  A deeper understanding for the cancer evolution and the heterogeneity at all levels is a potentially reachable goal that would enable control or even devise effective treatment of the fatal disease in the future. One of the obstacles slowing down  progress comes from the complex interactions among cancer cells, which leads to the formation of a complex cancer cell society, and to the emergence of unexpected phenomena\cite{tabassum2015tumorigenesis}.  This could serve as a platform for physicists to give detailed and quantitative understandings of the underlying mechanisms of complex systems using mathematical models. We have provided just one example to illustrate the possibility of applying physical models in studying important questions in cancer research. It is quite encouraging that a simple physical model  provides a unified description of different phenomena in pancreatic cancer and GBM in a quantitative manner. In addition, the model can also be used to explain the origin and maintenance of ITH in other cancers in which similar cell-cell interactions are present\cite{Cleary14,mateo2014sparc}. These ideas could be useful in microbial colonies, insect communities, human society, and other systems~\cite{dobata2013public,drescher2014solutions,kaul1999global}.   We expect that  similar physical models and ideas from very different fields\cite{altrock2015mathematics,kirkpatrick2015colloquium} could help us better understand the disease, and might also provide new ideas and methods for more efficacious cancer therapy.

\vspace{6mm}

\noindent {\bf Acknowledgements}\\
\noindent This work is supported by the National Science Foundation (PHY 17-08128 and CHE 16-32756), and the Collie-Welch Chair through the Welch Foundation (F-0019).

\bibliographystyle{vancouver}
\bibliography{cancerheterogeneitytexrevise_5}

\newpage
\clearpage

\newpage
\clearpage   

\newpage

\newpage

\newpage

\newpage

\newpage

\newpage

\begin{table}[]
\caption{Cancer types, estimated number of new cancer cases and deaths in United States, 2019\cite{siegel2019cancer}.}
\label{table1}
\begin{tabular}{lll}
\hline\hline
 &   Estimated New Cases   & Estimated Deaths \\
 \hline
 All sites &    1,762,450  & 606,880  \\
   \hline
 Tongue&     17,060& 3,020 \\
 Mouth&     14,310	&  2,740\\
 Pharynx&     17,870&  3,450\\
  Other oral cavity&    3,760 &1,650  \\
   \hline
 Esophagus&     17,650& 16,080 \\
 Stomach&     27,510& 11,140\\
Small intestine &    10,590 & 1,590 \\
Colon &    101,420 &  51,020\\
Rectum &    44,180 &  \\
 Anus, anal canal, and anorectum&  8,300   & 1,280 \\
 Liver and intrahepatic bile duct&  42,030   &31,780 \\
 Gallbladder and other biliary & 12,360    &  3,960\\
 Pancreas&   56,770  & 45,750 \\
 Other digestive organs&   7,220  & 2,860 \\
  \hline
 Larynx&    12,410&  3,760\\
 Lung and bronchus	&   228,150  & 142,670 \\
 Other respiratory organs	&5,880    & 1,080 \\
  \hline
Bones and joints	&   3,500  & 1,660 \\
 \hline
 Soft tissue (including heart)	&  12,750   &  5,270\\
  \hline
Skin (excluding basal and squamous)	&104,350	 &11,650\\
  \hline
 Breast	&271,270	 & 42,260 \\
 \hline
 Uterine cervix	&13,170		&4,250	 \\
 Uterine corpus	 &61,880		&	12,160	 \\
 Ovary	&22,530			&13,980    \\
 Vulva	&6,070		&1,280	\\
 Vagina and other genital, female	&5,350		&1,430	\\
 Prostate	&174,650		&31,620 \\
 Testis	&9,560	&410 \\
 Penis and other genital, male	&2,080	&410	 \\
 \hline
 Urinary bladder	&80,470	&17,670	 \\
 Kidney and renal pelvis	&73,820	&14,770 \\
 Ureter and other urinary organs	&3,930	&	980 \\
  \hline
 Eye and orbit	&3,360	&370 \\
 \hline
 Brain and other nervous system	&23,820	&17,760 \\
 \hline
Endocrine system	&54,740	&3,210	 \\
  \hline
 blood cancer     & 176,200    & 56,770 \\
  \hline
Other and unspecified primary sites &	31,480	&45,140 \\
  \hline
  \end{tabular}
\end{table}

\end{document}